# SPLAY AND TILT ENERGY OF BIPOLAR LIPID MEMBRANES

Timur R. Galimzyanov[*,†], Peter I. Kuzmin[*], Peter Pohl[‡], Sergey A. Akimov[*,†]

*Laboratory of Bioelectrochemistry, A.N. Frumkin Institute of Physical Chemistry and Electrochemistry, Russian Academy of Sciences, 31/4 Leninskiy prospekt, Moscow 119071, Russia; †Department of Theoretical Physics and Quantum Technologies, National University of Science and Technology "MISiS", 4 Leninskiy prospect, Moscow 119049, Russia; ‡Institute of Biophysics, Johannes Kepler University Linz, Gruberstr. 40-42, Linz, 4020, Austria.

## ABSTRACT

Archaea organisms are able to survive in extremely aggressive environment. It's thought that such resistance, at least, in part is sustained by unique properties of archaea membrane. The membrane consists of so called bolalipids, which has two polar heads joined by two hydrocarbon chains. Thus bolalipids can exist in two conformations: i) polar heads are located at different sides of bolalipid layer, so called, O-shape; ii) polar heads are located at the same side of the layer, so called, U-shape. Both polar heads and chains are chemically different from those for "conventional" lipids. In the present study we develop basis for theory of elasticity of bolalipid membranes. Deformations of splay, tilt and Gaussian curvature are considered. We show that energetic contributions of tilt deformation from two surfaces of bolalipid layer are additive, as well as Gaussian curvature, while splay deformations yield a cross-term. The presence of U-shapes is taken into account in terms of the layer spontaneous curvature. Estimation of tilt modulus and possible experiments allowing to measure splay moduli are described.

## INTRODUCTION

There are three kingdoms of life: bacteria, eukaryote and archaea (1). Archaea organisms often exist under extreme conditions, such as high pressure (~400 atm.), high temperatures (~100°C), high methane concentrations and very low or high environment acidity (2). As opposed to bacteria and eukaryote, the archaea cell membrane is formed by unique components, so-called bolalipids (bipolar lipids), which are believed to be responsible for the phenomenal stability of archaea organisms' membranes under extreme external conditions. "Conventional" lipids, that are characteristic for eukaryotic cells, consist of a polar head joint with two hydrocarbon chains. Under certain conditions, these lipids self-organize into bilayer structures (3). Bolalipids consist of two polar heads joined by two hydrocarbon chains. Such bipolar molecules form single layers in water (4). Bolalipid membranes are considered to be a promising material for various scientific and engineering applications (5–6), making the investigation of their distinctive thermodynamical properties very important.

Theoretical investigations of conventional lipids have been carried out in the framework of microscopic and macroscopic models. Microscopic models are represented by various molecular dynamic models (7) and analytical solutions of statistical mechanics equations (8). Macroscopic models are represented by elasticity theory that treats membranes as a continuum elastic medium. Here we have focused on the lipid membrane elasticity theories. The first elasticity theory for conventional lipid membranes was developed by Helfrich (9). Despite the simplicity of Helfrich's model, it was successfully utilized for theoretical investigations of membrane structures and membrane-associated phenomena (10–13). Another big step towards complete elasticity theory was work done by Hamm and Kozlov (14), in which the authors accounted for the bilayer intrinsic structure in the framework of its so-called tilt deformation. This theory is still widely used for the investigation of various membrane processes and phenomena, such as poration, fission, fusion, domain formation etc. (15–21). The theory-based models enable systematization of available experimental data and possess substantial predictive power. However, in view of bolalipids' structural features, the afore-mentioned elasticity theory cannot be directly applied to bolalipid membranes.

Bolalipids have been experimentally investigated for a long time (5). However, only little theoretical research has been carried out, and all of it was in the framework of microscopic models: by means of molecular dynamics (7, 22), and analytical solutions of equations from statistical mechanics (8, 23). A macroscopic elasticity theory for bolalipid membranes has not yet been developed.

Bolalipid molecules differ from the conventional lipid molecules, as they have two polar head-groups joined by two hydrophobic chains. They have two conformations: 1) So-called, O-shapes, in which polar heads are located at different sides of the membrane (Fig. 1 *a*); 2) So-called U-shapes, in which both polar heads are located at the same side of the membrane. (Fig. 1 *b*, *c*).

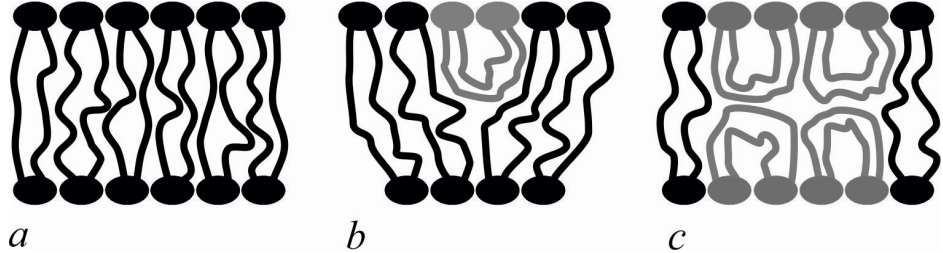

FIGURE 1. Possible bolalipid configuration in the membrane: *a* — O-shape; *b* — U-shape and O-shape mixture; *c* — U-shape forming bilayer structure.

In the general case, the bolalipid layer consists of both conformations. This was shown in NMR experiments, where the U-shape concentration was found to be about $\leq 10\%$ (4); and in numerical experiments (7) where the U-shape concentration was shown to vary from 0 to 60% depending on the particular experimental setup and the molecular properties. Therefore, the bolalipid membrane is quite a new object in comparison with conventional lipid membranes and demands a different approach in the elasticity formalism. Development of this approach is the main aim of the present work.

Firstly we derive a general expression for the energy surface density of bolalipid membranes that consist exclusively of O-shaped lipids. As a starting point we use the general elasticity theory of lipid membranes (14). Secondly, we consider U-shapes contribution to the elastic energy. Thirdly, we suggest possible experiments for defining elasticity moduli and others parameters of our model.

## STATEMENT OF THE PROBLEM

We treat the membrane as a continuous medium, which may be subjected to elastic deformations. For the aim of constructing an elasticity theory for bolalipid membranes, we assume that deformations are small and calculate their energy. In the first step we consider bolalipid membranes that only consist of O-shapes.

Similar to conventional lipids, deformations of the bolalipid layer may best be described in terms of shapes of two surfaces, located nearby bolalipid molecules polar heads and hydrocarbon tails joint at different sides of the layer. The surfaces are referred to as "dividing surfaces" (12). The shape of dividing surfaces is defined by vectors of unit normal **N** to them. Each surface is correlated with half of the membrane. That is, a third surface, the so called "midplane" is thought to divide the membrane. It is located somewhere between the two dividing surfaces. We discuss its exact position below in the text. For convenience, we will call the two halves of bolalipid membranes "monolayers". Since membranes only consist of O-shaped bolalipids that pierce through the membrane, monolayer deformations should be continuous across the membrane. The average orientation of bolalipids in each monolayer is characterized by the unit vector **n**, called "director". Thus, the bolalipid layer is characterized by the shape of two dividing surfaces and two vector fields of directors, defined at the corresponding dividing surfaces. In the membrane's unstrained state, dividing surfaces are parallel to each other, both unit normal **N** vectors and director vectors are collinear. Similar to membranes from conventional lipids, bolalipid membranes are considered both laterally liquid and locally volumetrically incompressible.

In the first part of this work, we followed the algorithm described in (14). For convenience, we provide the necessary basic equations of this paper below without excessive mathematical details. Eq. 1 is the general expression for the elastic energy of laterally liquid media, written up to the second-order term:

$$dF = dV\left[\sigma_L \varepsilon + \frac{1}{2}\lambda_L \varepsilon^2 + \frac{1}{2}(4\lambda_T)u_{z\alpha}u^{z\alpha}\right], \tag{1}$$

where $u_i$ denotes the components of the displacement vector $\mathbf{u} = \mathbf{r} - \mathbf{r}_0$, $\mathbf{r}_0$ is the radius-vector of the volume element in the non-deformed state, $\mathbf{r}$ is the radius-vector of the volume element in the deformed state, $u_{ij}$ are deformation tensor components, defined by the components of displacement vector $\mathbf{u}$: $u_{ij} = \frac{1}{2}(\nabla_i u_j + \nabla_j u_i) + ((\nabla \mathbf{u})^T \nabla \mathbf{u})_{ji}$; $\sigma_L$, $\lambda_T$ are elastic moduli. For further convenience, the relative lateral expansion of the volume element, $\varepsilon$, is written explicitly rather than through the displacement vector. The volumetric incompressibility condition allowed us to connect $\varepsilon$ with deformation vector components: $(1+\varepsilon)(1+\nabla_z u_z) = 1$, or up to the second order terms of $\varepsilon$: $\nabla_z u_z = -\varepsilon + \varepsilon^2 + ...$. Eq. 1 is given in the lab coordinate system, where $z$ axis is directed perpendicular to $Oxy$ plane of an unstrained membrane.

The final expression for the surface density of monolayer free energy in (14) is written in terms of splay and tilt deformations. Tilt deformation is characterized by the so-called tilt-vector, defined as **t** = **n**/(**nN**) – **N**. Splay deformation is characterized by the effective curvature $\tilde{J}$, defined by traces of surface curvature tensor $b^{\alpha}_{\beta}$ and a variation of tilt-vector: $\tilde{J} = b^{\alpha}_{\alpha} - \nabla_{\alpha} t^{\alpha}$. The curvature tensor $b^{\alpha}_{\beta}$ is defined by the equation that connects the derivatives of the unit normal vector in the lab coordinate system $\{x_i\}$ and the local tangential basis on the dividing surface $\{\mathbf{r}_i\}$: $\dfrac{\partial \mathbf{N}}{\partial x_{\alpha}} = -b^{\beta}_{\alpha} \mathbf{r}_{\beta}$.

In (14) Eq. 1 is applied to a small area of a lipid monolayer patch. The deformation of the small patch volume is a linear function of the distance between the volume element and the dividing surface. In addition, this function is linear in tilt and splay deformations. Parameterization functions are unambiguously defined by director and unit normal vectors, which are set at a so-called "neutral surface". It is a surface within the monolayer where cross-terms between splay and compression/stretching vanish. According to (12) this surface is located close to the region where the polar lipid heads join with hydrocarbon tails.

The final expression for the energy surface density, obtained from Eq. 1, is integrated over monolayer thickness and formulated in terms of tilt and splay deformations. Tilt and splay deformations are found to be independent of each other. The given outline is projected to each monolayer of the bolalipid membrane.

## SOLUTION OF THE PROBLEM

**Tilt deformation of bolalipid membrane.** In tilt deformation, the director deviates from the normal of each dividing surface (Fig. 2 *b*). The values corresponding to different dividing surfaces are denoted by indices «1» (also called "bottom") and «2» (also called "upper"); at the unstrained membrane, the *z* axis is directed from the bottom towards the upper dividing surface.

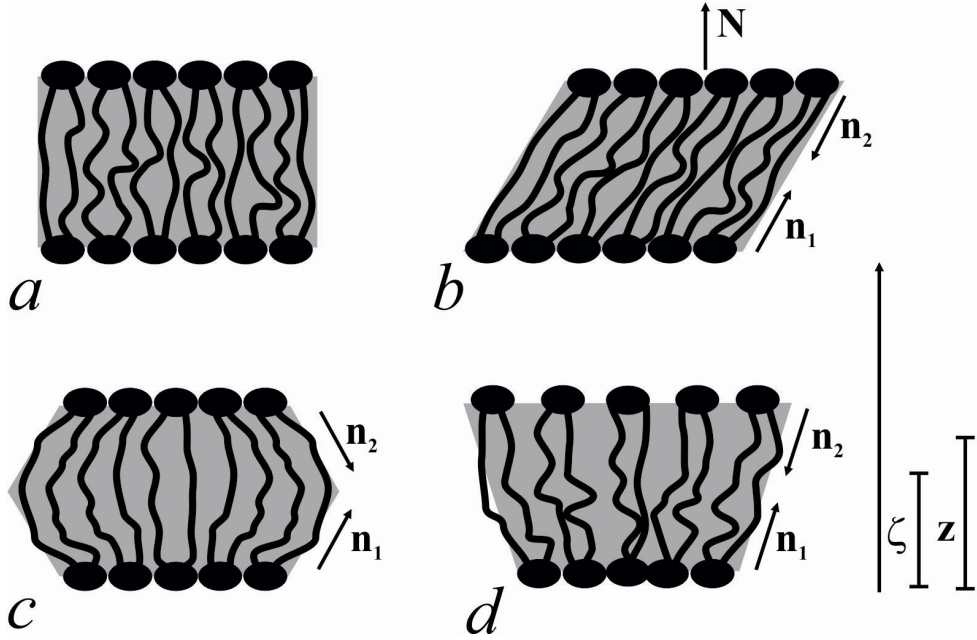

FIGURE 2. Deformations of bolalipid membrane. *a* — Unstrained membrane patch; *b* — Uniform tilt deformation; *c* — Symmetric splay deformation; *d* — Asymmetric splay deformation. Bars show different scales of $\zeta$ and *z*-axes of local tangential and lab coordinate system, respectively.

Tilt deformation reduces to the following dependence of the deformation vector on *z*-coordinate (14):

$$\mathbf{u} = \begin{cases} \mathbf{t}_1 \cdot z, & z < h_m, \\ \mathbf{t}_2 \cdot (2h_0 - z), & z > h_m, \end{cases} \quad (2)$$

where $h_m$ is the distance between midplane and bottom dividing surface; $2h_0$ is the equilibrium membrane thickness; **t** is so-called tilt vector. For small deformations, the tilt-vector is defined as the difference between the director and the dividing surface unit normal vector, $\mathbf{t} = \mathbf{n} - \mathbf{N}$. The condition that deformation should be continuous everywhere, in particular, at the midplane, i.e. $\mathbf{u}(h_m - 0) = \mathbf{u}(h_m + 0)$ leads to the small modification of Eq. 2:

$$\mathbf{u} = \begin{cases} \mathbf{t}_1 \cdot z & z < h_m \\ \mathbf{t}_2 \cdot (h_m - z) + \vec{t}_1 h_m & z > h_m \end{cases} \quad (3),$$

The only nonzero deformation tensor components $u_{z\alpha}$ take the following form: $u_{z\alpha} = \frac{1}{2}\left[t_{1,\alpha} \cdot \theta(h_m - z) - t_{2,\alpha} \cdot \theta(z - h_m)\right]$ ($\alpha$ = x, y), where $\theta(z)$ is the Heaviside step function, defined as:

$$\theta(z) = \begin{cases} 0, & z < 0, \\ 1, & z \geq 0. \end{cases} \tag{4}$$

Inserting these deformation tensor components to Eq. 1, we obtain the contribution of tilt deformation to the elastic energy:

$$dF_t = \frac{\lambda_T(z)}{2}\left(t_1^2 \cdot \theta(h_m - z) + t_2^2 \cdot \theta(z - h_m)\right),$$
$$F_t = \frac{1}{2}t_1^2 \cdot \int_0^{h_m} \lambda_T(z)dz + t_2^2 \cdot \int_{h_m}^{2h_0} \lambda_T(z)dz. \tag{5}$$

The midplane position is defined by the distance $h_m$ between the bottom dividing surface and the midplane. Commonly it depends on various factors and may vary in lateral direction. Obviously, in cases where the monolayer properties and deformations are symmetric, the midplane is located in the middle of the bolalipid layer. For the limiting case of zero symmetric deformations, we conclude that $h_m = h_0$ in the unstrained membrane. Upon deformations, the midplane position deviation is characterized by the value $(h_m - h_0)/h_0$, which is of the same (or higher) infinitesimal order as the deformations. So, accounting for the deviations of $h_m$ from its equilibrium value $h_m = h_0$ in Eq. 5 leads to higher than second orders terms. Thus, with regard to the required accuracy, we assume $h_m = h_0$. All of the aforesaid leads to the expression for the tilt term of the elastic energy for the bolalipid membrane:

$$F_t = \frac{1}{2}K_t t_1^2 + \frac{1}{2}K_t t_2^2,$$
$$K_t = \int_0^{h_0} \lambda_T(z)dz = \int_{h_0}^{2h_0} \lambda_T(z)dz. \tag{6}$$

Thus, we have derived the expression for the surface energy density of the uniformly tilted membrane without any cross-terms on tilts. The reason for the absence of a term proportional to $t_1 t_2$ is not straightforward and needs some explanations. Cross-terms are connected with contributions from the average curvature of lipid hydrocarbon tails, which in our approach appears to be negligible in comparison with the energy of tilt and splay deformations. These results match experimental data (4, 7), according to which a substantial U-shape concentration is found in the bolalipid membrane. This means that energy of even such a significantly curved hydrocarbon chain is comparable with the characteristic energy of thermal fluctuations, $k_B T$, and with the energy of elastic deformations. Consequently, small average curvatures of the chain would lead to significantly smaller energetic costs.

**Splay deformation of bolalipid membrane.** In this part we consider uniformly curved membranes (Fig. 2 *c*, *d*). As was noted above, splay deformation is independent on tilt deformations, i.e. uniform splay does not lead to shear of volume elements and $u_{za} = 0$. Splay contributions to the elastic energy are due to the stretching the hydrocarbon region ($\varepsilon \neq 0$). Leaving aside differential geometry details provided in the paper (14), we note that a small deformed patch may be treated in terms of a curvilinear trapezium in order to calculate splay contributions to the elastic energy. For conventional lipids, stretching of monolayer volume elements, $\varepsilon$, is proportional to the mean and Gaussian curvatures values, $J$ and $K$: $\varepsilon = \zeta J + \zeta^2 K$, where $\zeta$ is the distance between the bottom dividing surface and the volume element in the tangential coordinate system, accounting for change in the membrane thickness due to the volumetric incompressibility condition (Fig. 2 *d*). Monolayer curvature is defined at the neutral surface, which is not stretched under splay deformation. For further calculation, we should note that Gaussian curvature value $K$ is of the second order of smallness, while the mean curvature value $J$ is of the first order (14).

Bolalipid membrane deformations are parameterized by two pairs of curvatures: mean and Gaussian ones, relevant to the bottom and upper dividing surfaces — $J_1$, $\kappa_1$ and $J_2$, $\kappa_2$ respectively. In this case, volume element stretching takes the following form:

$$\varepsilon = \left(-\zeta J_1 + \zeta^2 \kappa_1\right)\theta\left(h_s - \zeta\right) - \left(\left(d - \zeta\right)J_2 + \left(d - \zeta\right)^2 \kappa_2\right)\theta\left(-h_s + \zeta\right), \tag{7}$$

where $d$ is the thickness of the curved membrane, $h_s$ is the position of midplane in the tangential coordinate system. Splay values for symmetric deformations (Fig. 2 *c*) are defined at the neutral surfaces as in the case of conventional lipid membranes. In contrast to membranes from conventional lipids, we cannot identify two neutral surfaces in bolalipid membranes that are subjected to asymmetric splay (Fig. 2 *d*). For example, formation of a closed vesicle from a bolalipid membrane should result in area changes of the dividing surfaces of opposite monolayers (Fig. 2 *d*). However, there is a sole neutral surface that is located around the middle of the bolalipid layer.

The proximity of the neutral surface to the region of polar heads in conventional lipid membranes indicates that this region has a substantially higher stretching modulus than the region of hydrophobic tails. Supposing that this is also valid for bolalipids, we find that stretching occurs only when the whole membrane experiences splay deformations, i.e. when the curvature difference between bottom and upper dividing surfaces is nonzero (Fig. 2 *d*). Moreover, within the framework of linear theory (Hooke's law), two additional assumptions are satisfied: (i) stretching depends only on the curvature so that the dividing surfaces are stretched in proportion to their curvature differences; (ii) stretching energy is equally distributed between the dividing surfaces.

For arbitrary stretching of dividing surfaces, Eq. 7 can be written in the form:

$$\varepsilon = \left(\varepsilon_1 - \zeta J_1 + \zeta^2 \kappa_1\right)\theta(h_s - \zeta) + \left(\varepsilon_2 - (d-\zeta)J_2 + (d-\zeta)^2 \kappa_2\right)\theta(-h_s + \zeta) \qquad (8)$$

where the indices of $\varepsilon_1$, $\varepsilon_2$ denote the bottom and upper dividing surfaces, respectively. The stretching field must be continuous so that $\varepsilon(h_s - 0) = \varepsilon(h_s + 0)$ should be added to Eq. 8. Assumptions (i) and (ii) lead to the following equation: $\varepsilon_1 = -\varepsilon_2 = \varepsilon_0 = \dfrac{J_1 - J_2}{2}h_0 - \dfrac{\kappa_1 - \kappa_2}{2}h_0^2$.

For further calculations, the transition from the local tangential coordinate system ($\zeta$ coordinate) to the lab one (*z* coordinate) should be made by means of volumetric incompressibility conditions: $A_0 z = A_0 \int_0^\zeta (1 + \varepsilon(\zeta'))d\zeta'$, where $A_0$ is the area of the membrane patch. So, the relation between $\zeta$ and *z* is the following:

$$\begin{cases} z = (1+\varepsilon_0)\zeta - \dfrac{1}{2}J_1\zeta^2 + \dfrac{1}{3}\kappa_1\zeta^3, & \zeta < d/2 \\ 2h_0 - z = (1-\varepsilon_0)(d-\zeta) - \dfrac{1}{2}(d-\zeta)^2 J_2 + \dfrac{1}{3}\kappa_1(d-\zeta)^3, & \zeta > d/2 \end{cases} \qquad (9)$$

Expressing $\zeta$ in terms of $z$ and substituting the result to Eq. 7, we obtain the stretching $\varepsilon(z)$ as a function of distance $z$ between the volume element and the bottom dividing surface:

$$\varepsilon = \begin{cases} \dfrac{z(h_0 - z) J_1^2}{2} - \dfrac{h_0 z}{2} J_1 J_2 + (h_0/2 - z) J_1 - \dfrac{h_0}{2} J_2 + \\ + (z^2 - h_0^2/2) \kappa_1 + \dfrac{h_0^2}{2} \kappa_2, \quad z < h_s, \\ \dfrac{z(z - h_0) J_2^2}{2} - \dfrac{h_0(2h_0 - z)}{2} J_1 J_2 + (-3h_0/2 + z) J_2 - \dfrac{h_0}{2} J_1 + \\ + \left((2h_0 - z)^2 - h_0^2/2\right) \kappa_2 + \dfrac{h_0^2}{2} \kappa_1, \quad z > h_s. \end{cases} \quad (10)$$

Substituting Eq. 10 into Eq. 1, and recalling that $u_{za} = 0$ for pure splay, we obtain the splay contribution to the elastic energy:

$$dF_J = \frac{1}{4} B_s (J_1 + J_2 - J_{ss})^2 + \frac{1}{4} B_d (J_1 - J_2 - J_{ds})^2 + K_G (\kappa_1 + \kappa_2) \quad (11)$$

where elastic moduli are defined as follows: $B_s = \int_0^{h_0} (\lambda - \sigma) z^2 dz$, $B_d = \int_0^{h_0} ((\lambda - \sigma) z(z - d)) dz + h_0^2 \int_0^{h_0} \lambda dz$, $K_G = \int_0^{h_0} \sigma(z^2 - h_0^2/2) dz$. The spontaneous curvatures $J_{ss}$ and $J_{ds}$ are determined by the expressions: $B_s J_{ss} = \tau_s$, $B_d J_{ds} = \tau_d$, where $\tau_s = \int_0^{h_0} \sigma z dz + \int_{h_0}^{d} \sigma \cdot (d - z) dz$, $\tau_d = \int_0^{d} \sigma(z - h_0) dz$. The elastic modulus $B_d$ corresponds to the splay modulus of the whole membrane (monolayer curvatures with equal absolute values and opposite signs); $B_s$ corresponds to the intrinsic membrane splay that acts to preserve a flat membrane shape on average (the curvatures of the monolayers are equal both in absolute value and sign). $J_{ss}$ and $J_{sd}$ are similar to the spontaneous curvatures of conventional lipid membranes. They can be considered as the sum and the difference of monolayers' spontaneous curvatures. $K_G$ is the Gaussian curvature modulus.

The expression for the total energy of an arbitrarily deformed small patch of a bolalipid membrane is then:

$$dF_J = \frac{B_s}{4}(J_1 + J_2 - J_{ss})^2 + \frac{B_d}{4}(J_1 - J_2 - J_{ds})^2 + \frac{K_t}{2}(t_1^2 + t_2^2) + K_G(\kappa_1 + \kappa_2) \quad (12)$$

Thus, we obtained the surface energy density of elastic deformations including both mean and Gaussian curvatures and tilt deformation. In contrast to the corresponding expressions for the conventional lipid bilayer, the cross-term for the curvatures of opposite monolayers exists, yet the cross-term for the tilts of opposite monolayers is absent. The Gaussian curvature cross-term does not exist, since it would exceed the accuracy of the model.

**Spontaneous curvature.** Expression Eq. 12 is valid for membranes from O-shaped bolalipids. Membranes from U-shaped molecules are described by the elastic energy density expression for conventional lipids, since its two monolayers have independent deformation characteristics. However, the spontaneous curvature of a monolayer from U-shaped bolalipids is likely to be positive, since both polar head groups are located at the same side of the monolayer (Fig. 3 *a*).

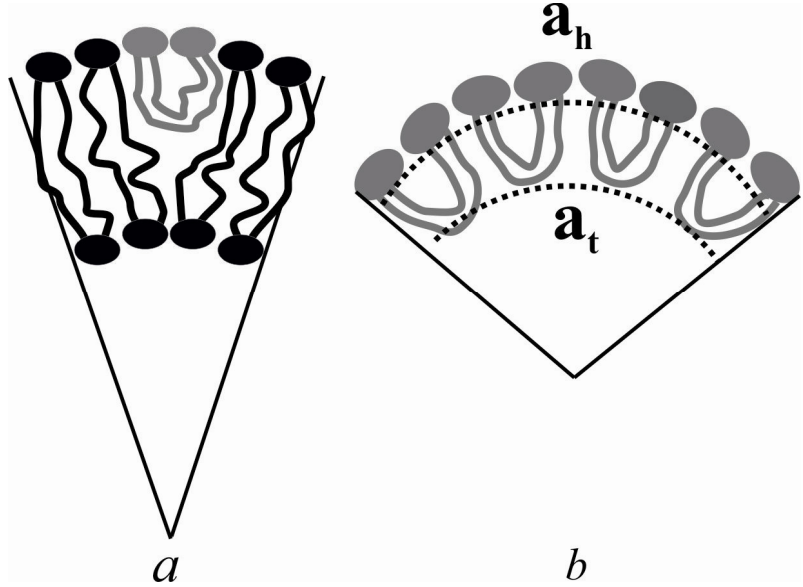

FIGURE 3. *a* — U-shaped lipids should induce a spontaneous curvature in bolalipid membranes. *b* — toy-model of a bolalipid monolayer from U-shapes.

Consequently, the elastic energy functional for a mixed layer from U- and O-shaped bolalipids is solved by attributing a spontaneous curvature to U-shape configurations. In the framework of linear theory, this spontaneous curvature is a linear function of the relative concentrations of the individual components. Assuming zero spontaneous curvature of O-shapes, the spontaneous curvature of a layer that is mixed from O- and U-shapes is equal to:

$J_s = xJ_{sU}$, where $x$ is the U-shape concentration in the membrane monolayer. In such notations $J_{ss}$ and $J_{ds}$ adopt the form:

$$J_{ss} = J_{sU} \cdot (x_1 + x_2)$$
$$J_{ds} = J_{sU} \cdot (x_1 - x_2)$$
(13)

where $x_1$, $x_2$ are the concentrations of U-shapes in the bottom and upper monolayers, respectively. Accordingly, the energy surface density of elastic deformations of bolalipid membranes is given as:

$$f = \frac{1}{4}B_s\left(J_1 + J_2 - J_{sU} \cdot (x_1 + x_2)\right)^2 + \frac{1}{4}B_d\left(J_1 - J_2 - J_{sU} \cdot (x_1 - x_2)\right)^2 + \frac{K_t}{2}\left(t_1^2 + t_2^2\right) + K_G(\kappa_1 + \kappa_2)$$
(14)

This expression ignores the entropic contribution of mixing U-shapes with O-shapes. Any lateral inhomogeneity of U-shapes may favor membrane deformations that are laterally non-uniform.

**Elastic modulus of tilt**. The elastic moduli $B_s$, $B_d$, $K_t$, $K_G$ (compare Eq. 14) should be measured experimentally, calculated from the microscopic models, or otherwise assessed. Simple calculations show that conventional lipids' tilt modulus should be close to the surface tension of the oil-water interface (14), which was confirmed by experimental data (24). If the same considerations are extended to bolalipid membranes, the bolalipid tilt modulus is estimated to be equal to the tilt modulus of conventional lipids, i.e. $K_t \sim 50$ dyn/cm.

All other elastic moduli depend on lipid structures and properties, thereby precluding this kind of simple estimations. They should be measured experimentally. However, the experimental definition of the Gaussian curvature modulus is very complicated even in the case of conventional lipids. At the same time, the Gaussian curvature term only needs to be accounted for in a narrow and peculiar set of problems, in which topological changes take place. We focus on the description of possible methods of splay moduli determination.

**Splay modulus $B_d$**. Measurements of lipid bilayer splay modulus are commonly based on monitoring membrane area changes that are associated with shape fluctuations. The relationship between a vesicle's fluctuational extra area with the splay modulus and its surface tension was derived theoretically and tested experimentally (25–28).

Giant unilamellar vesicles (GUVs) with a diameter of about 10 μm are well suited for the purpose because the average curvature is small. $J_1$ and $J_2$ have different signs. Thus, in case of small curvature, $J_1 + J_2$ is much smaller than $J_1 - J_2$. This means that only the $B_d$ modulus can be determined by such an experiment. Since bolalipids form GUVs [Dr. O. V. Batishchev, personal communication] the experiment is feasible. The energetic contribution of the Gaussian curvature is constant because the system's topology does not change during the experiment (Gauss-Bonnet theorem).

**Splay modulus $B_s$.** Conductivity measurements of lipid nanotubes that are pulled from the membrane represent an alternative method for the determination of elastic properties. They reveal the nanotube radius $R = 1/J$. $R$ depends both on splay modulus and membrane lateral tension (29–30). $J_1 + J_2$ cannot be assumed to be small because $R$ is comparable with membrane thickness. Moreover, the U-shaped bolalipids are likely to redistribute laterally. Due to the cylindrical symmetry of the nanotube, tilt deformations do not appear. In addition, Gaussian curvature does not contribute to the energy associated with changes in $R$.

The linear density of elastic energy of a cylindrical tube that is subjected to external lateral tension $\sigma$ is the following:

$$F = \frac{2\pi}{J}\left[\frac{1}{4}B_s\left(J_1 + J_2 - J_{ss}\right)^2 + \frac{1}{4}B_d\left(J_1 - J_2 - J_{sd}\right)^2\right] + \sigma\left(\frac{2\pi}{J_1} + \frac{2\pi}{J_2}\right), \quad (15)$$

where $J_1 = (1/J + h)^{-1}$, $J_2 = -(1/J - h)^{-1}$. Indices "1" and "2" correspond to external and internal monolayers, respectively. We define $R$ at the membrane midplane; $h$ is monolayer thickness, $J_{ss}$ and $J_{sd}$ are spontaneous curvatures (Eq. 13). The energy density is multiplied by the area of the non-deformed state (31), which with necessary accuracy is equal to the area of the nanotube midplane. $F$ (Eq. 15) should be minimized with respect to $J$ and the concentration of U-shapes, which will result in equilibrium (measured) nanotube radius as a function of lateral tension, $\sigma$. These parameters can be obtained independently by varying the lateral tension via application of transmembrane voltage (29).

For conventional lipids, elastic moduli are much greater than the characteristic energy of thermal fluctuations, $k_BT$. For instance, the characteristic splay modulus value is about 10 $k_BT$ (28), while the characteristic entropic energy is 1 $k_BT$. We thus may assume that lateral distribution of U-shapes is governed by membrane elastic energy. Formation of nanotubes is much faster than the lateral redistribution of membrane components with non-zero spontaneous curvature (U-shape) (32). Consequently, immediately after formation, the U-shape concentration in the internal and the external monolayers of the nanotube are the same as in flat membranes. Subsequently, the nanotube radius relaxes due to the lateral redistribution of U-shapes. The relaxation is governed by the minimization of elastic energy. Its characteristic time amounts to about 1 s for conventional lipids (dioleoylphosphatidylethanolamine, DOPE) (32).

Thus, immediately after nanotube formation its composition is symmetric, and $J_{sd} = 0$. Minimizing $F$ (Eq. 15) with respect to nanotube curvature, we obtain the following expression for $B_s$:

$$B_s = \frac{B_d\left(\left((h/R)^2+3\right)\left(1-(h/R)^2\right)hJ_{ss} - \frac{3(h/R)^4+8(h/R)^2+1}{h/R}\right) + \frac{\left(1-(h/R)^2\right)^4 R^3}{(h/R)R'}}{6h/R\left((h/R)^2+1\right)},$$

(16)

$x$ is the concentration of U-shapes in the flat membrane, $R'$ is derivative of the nanotube radius with respect to lateral tension $\sigma$. The expression can be simplified if it is considered that: (i) $J_{ss} = 2xJ_{sU}$ (Eq. 13). For a small ratio $h/R$ we yield:

$$B_s \approx -\frac{1}{6(h/R)^2}\left(B_d - \frac{R^3}{R'}\right)$$

(17)

Subsequent to the lateral redistribution of U-shapes, the nanotube state can be obtained by substituting the spontaneous curvature given by Eq. 13 into Eq. 15 and minimizing the energy with respect to the concentration of U-shapes and $R$. Energy minimization demands the absence of configurations with negative curvature in the internal monolayer of the nanotube because $J_s$ is always positive. It yields:

$$\frac{2\sigma}{J^2} = \frac{B_s B_d (1+hJ)}{(B_s + B_d)(1-hJ)^3} \tag{18}$$

Derivation of Eq. 18 with respect to $\sigma$ results in:

$$\frac{B_s B_d}{B_s + B_d} = \frac{R^3}{R'} \frac{(1-h/R)^4}{1+2h/R} \tag{19}$$

where $R'$ and $R$ are measurable parameters. Thus, Eq. 19 gives the combination of splay moduli $\frac{B_s B_d}{B_s + B_d}$. Knowledge of $B_d$ from experiments with GUVs (see above), allows us to determine the value of elastic modulus $B_s$. Thus, a combination of two methods provides both splay moduli of bolalipid membranes.

## DISCUSSION

We have obtained a general expression for the surface energy density of elastic deformations for bolalipid membranes, consisting of two types of molecules: O-shapes and U-shapes. The energy includes cross terms for curvatures of opposite monolayers as well as for curvatures and U-shape concentrations. Tilt cross-terms are absent because they are determined by the average hydrocarbon chain bending, which is negligible in the framework of the approach used.

Experiments for the determination of two splay moduli of bolalipid membranes were proposed. The moduli of elasticity and spontaneous curvature of U-shape monolayers have to be assessed by theoretical considerations. In zero approximation the tilt modulus can be taken to be equal to conventional lipids' tilt modulus. The spontaneous curvature of a monolayer from U-shapes can be estimated using a toy-model (Fig. 3 *b*). Therefore we assume that the monolayer spontaneously adopts the shape of a spherical segment of radius $R_s$, i.e. of curvature $2/R_s$. $R_s$ is found from the area per U-shaped molecule both in the head-group ($a_h$) and in the tail regions ($a_t$) of the monolayer:

$$J_{sU} = \frac{1}{h_0} \frac{a_h - a_t}{\left(a_h + \sqrt{a_h a_t}\right)}, \tag{20}$$

where $h_0$ is the equilibrium thickness of the monolayer from U-shapes, which can be taken to be equal to half of the thickness of membranes from O-shapes. In the simplest case the spontaneous curvature of a layer from O-shapes is equal to zero and, thus, that the average area of a O-shape lipid in the polar head region is equal to the average lipid area in the middle of the tail region. This allows us to estimate that the area $a_h$ of two polar heads of U-shapes is roughly twice as large as the average tail area $a_t$. Substituting this into Eq. 20 results in a simple expression for spontaneous curvature of a monolayer from U-shapes: $J_{sU} = \frac{1}{2 + \sqrt{2}} \frac{1}{h_0} \approx \frac{1}{3 h_0}$.

We have not yet considered the dependence of the elastic moduli on U-shape concentration. Since pure U-shape membranes are equivalent to conventional lipid membranes, the curvature cross-terms should vanish, i.e. $B_d = B_s$. The energy of pure O-shape membranes has curvature cross-terms and $B_d \neq B_s$. Thus, the dependence of the elastic moduli on the concentration of U-shapes should be taken into account for systems with large amounts of U-shaped molecules.

## AKNOWLEDGEMENTS

The work was supported by Government of Russian Federation through Goszadanie research project grant 3.2007.2014/K, by the Ministry of Education and Science of Russian Federation in the framework of Increase Competitiveness Program of MISiS. We gratefully acknowledge the financial support of the Russian Foundation for Basic Research (grants ## 13-04-40325, 13-04-40327); Program of Presidium of Russian Academy of Sciences "Molecular and Cell Biology".